\documentclass[11pt]{IEEEtran}
\begin{document}

\newtheorem{Lemma}{\it Lemma}
\newtheorem{Theorem}{\it Theorem}
\newtheorem{Corollary}{\it Corollary}
\newtheorem{Definition}{\it Definition}
\newtheorem{Example}{\it Example}

\newpage
\setcounter{page}{8}

\title{Comment on ``A Random Quantum Key Distribution Achieved by Using Bell States''
\thanks{This work is supported by the China National Natural Science Foundation
Grant No. 60373059, the National Laboratory for Modern
Communications Science Foundation of China Grant No.
51436020103DZ4001, National Research Foundation for the Doctoral
Program of Higher Education of China Grant No. 20040013007.}}

\author{\small Sujuan Qin$^{1,2}$ \quad Qiaoyan Wen$^{1}$ \quad Fuchen Zhu$^{3}$\\
        (1. School of Science, Beijing University of Posts and Telecommunications,
        Beijing, 100876, P. R. China) \\
        (2. State Key Laboratory of Integrated Services Network, Xidian University,
        Xi'an, 710071, P. R. China)\\
        (3. National Laboratory for Modern Communication, P.O.Box 810, Chengdu, 610041, P. R. China) \\
        \vspace{4pt} E-mail: qsujuan2003@163.com}
\maketitle

\newpage
\setcounter{page}{1}

\begin{abstract}
In a recent paper [J. Opt. B: Quantum Semiclass. Opt. 5 (2003)
155-157], a quantum key distribution scheme based on entanglement
swapping was proposed, which exhibited two improvements over the
previous protocols. In this Comment, it is shown that the scheme
has no properties as been discussed.

\vspace{6pt} \hspace{1pt}\noindent{\bf Keywords:}\quad Quantum key
distribution, Bell operator, entanglement swapping
\end{abstract}

\vspace{22pt} Chong Li \textit{et al}. presented a quantum key
distribution (QKD) scheme based on entanglement swapping [1]. Two
improvements over the previous protocols were exhibited:

(1) If the communicators share enough known entangled pairs before
the key supply, the sender (Alice) need not send any particle to
the receiver (Bob).

(2) The efficiency can approach four-bit secret communication per
two entangled pairs (in the BB84 protocol, only one bit per pair
of particles is achieved).

The aim of this comment, however, is to point out that there are
no such two advantages in Ref. [1].

As to advantage (1), after finishing the communication, Alice and
Bob no longer share any entangled pair. That is these entangled
pairs shared before cannot be reused as in [2,3]. So the
communicators must share enough known entangled pairs before the
key supply in every communication. i.e. sharing entangled pairs is
the premise. So for a complete protocol, we must consider how to
share entangled pairs first. There are many ways to accomplish
this. For example, Alice (Bob) prepares a sequence of EPR pairs
and then sends one particle of each pair to Bob (Alice); or a
trusted third party prepares a sequence of EPR pairs and then
sends each particle from each pair to Alice and Bob respectively;
or Alice and Bob share EPR pairs in other ways. However, particles
are sent in all these methods. Many protocols [4,5] need not send
any particle if communicators share enough known entangled pairs
initially. So we have no reason to confess that the advantage (1)
stands.

As to advantage (2), the following equality holds according to the
feather of entangle swapping (For convenience, we use the same
notation as in Ref. [1]).
$$
\begin{array}{rr}
|\Phi\rangle_{ABAB}^{1234}=\frac{1}{2}\{|\phi^+\rangle_{AB}^{13}|\psi^+\rangle_{AB}^{24}+|\phi^-\rangle_{AB}^{13}|\psi^-\rangle_{AB}^{24}\\
+|\psi^+\rangle_{AB}^{13}|\phi^+\rangle_{AB}^{24}+|\psi^-\rangle_{AB}^{13}|\phi^-\rangle_{AB}^{24}\}
\end{array}
$$
In the Ref. [1], we get 4 bit keys, 0010,1101,1000 or 0111.
However, from the Bell operator measurement results of Alice, we
can determine the results of Bob. That is to say, the results
between Alice and Bob have strict relevance. In light of the
information theory, we get the entropy of information:
$$H=-\sum_{i=1}^4p_ilogp_i=-\sum_{i=1}^4\frac{1}{4}log\frac{1}{4}=2bit$$
So the efficiency can only approach two bit keys per two entangled
pairs.

According to the above analysis, we show that there are no such
two advantages in Ref. [1].

\end{document}